\documentclass{IEEEtran4PSCC}

\usepackage[cmex10]{amsmath,mathtools}
\usepackage{bbm}
\usepackage[pdftex]{graphicx}
\usepackage{amsfonts}
\usepackage{tikz}
\usetikzlibrary{shapes,arrows}
\usetikzlibrary{arrows,decorations.pathmorphing,backgrounds,fit,positioning,shapes.symbols,chains, shapes,decorations}
\usepackage{hyperref}
\usetikzlibrary{arrows.meta}
\usetikzlibrary{shapes,arrows}
\usetikzlibrary{automata, positioning}
\usepackage{algorithm}
\usepackage{algpseudocode}
\usepackage{cite}
\usepackage{csquotes}
\makeatletter
\let\old@ps@headings\ps@headings
\let\old@ps@IEEEtitlepagestyle\ps@IEEEtitlepagestyle
\def\psccfooter#1{%
    \def\ps@headings{%
        \old@ps@headings%
        \def\@oddfoot{\strut\hfill#1\hfill\strut}%
        \def\@evenfoot{\strut\hfill#1\hfill\strut}%
    }%
    \def\ps@IEEEtitlepagestyle{%
        \old@ps@IEEEtitlepagestyle%
        \def\@oddfoot{\strut\hfill#1\hfill\strut}%
        \def\@evenfoot{\strut\hfill#1\hfill\strut}%
    }%
    \ps@headings%
}
\makeatother

\psccfooter{%
        \parbox{\textwidth}{\hrulefill \\ \small{21st Power Systems Computation Conference} \hfill \begin{minipage}{0.2\textwidth}\centering \vspace*{4pt} \includegraphics[scale=0.06]{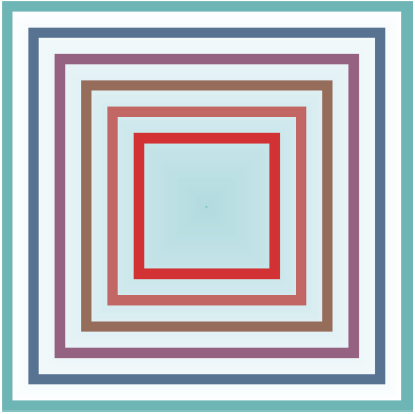}\\\small{PSCC 2020} \end{minipage} \hfill \small{Porto, Portugal --- June 29 -- July 3, 2020}}%
}

\begin{document}
\title{Data-Driven Learning and Load Ensemble Control}

 \author{\IEEEauthorblockN{Ali Hassan,
 Deepjyoti Deka,
 Michael Chertkov and Yury Dvorkin}
 }

\maketitle

\begin{abstract}
Demand response (DR) programs aim to  engage distributed small-scale flexible loads, such as thermostatically controllable loads (TCLs), to provide various grid support services. Linearly Solvable Markov Decision Process (LS-MDP), a variant of the traditional MDP, is used to model aggregated TCLs. Then, a model-free reinforcement learning technique called Z-learning is applied to learn  the value function and derive the optimal policy for the DR aggregator to control TCLs. The learning process is robust against  uncertainty that arises from estimating the passive dynamics of the aggregated TCLs. The efficiency of this data-driven learning is demonstrated through simulations on Heating, Cooling \& Ventilation (HVAC) units in a testbed neighborhood of residential houses.
\end{abstract}


\section{Introduction}
Distribution grids are undergoing a rapid transition due to the massive deployment of distributed energy resources (DERs), e.g., PV arrays, electric vehicles, and energy storage units. The main factors fueling this expansion include significant decreases in the capital costs  of DER technologies and incentives for DER installations offered by local electric power utilities, as well as by local and state authorities. For example, the state of California aims to reduce greenhouse gas emissions (GHG) by 40\% below its 1990 levels in 2030 by means of increasing the share of electricity produced by renewable generation to 50\%, doubling energy efficiency targets, and encouraging widespread transportation electrification \cite{california_target}. Similarly, the state of NY set a target of zero-carbon power sector by 2040, along with the goal of reducing the 1990 levels of GHG  emissions by 85\% in 2050 \cite{NY_target}. On the other hand, the presence of DERs in  distribution grids also imposes additional operational  challenges, e.g. bidirectional power flows, voltage fluctuations, and, as a result, additional wear-and-tear on electric power equipment. Dealing with such challenges is crucial to ensure economic and reliable distribution grid operations and necessitates more flexibility. Demand Response (DR) is one way to provide this additional flexibility, which enrolls controllable loads in residential and commercial buildings to provide a broad range of distribution-level ancillary services (e.g. energy arbitrage, peak shaving, balancing regulation, congestion relief, capacity deferral, voltage  support, \cite{aggregated_tcl_callaway}). Our efforts to explore this source of flexibility is motivated by the recent statistics that the U.S. building sector claims about 40\% of the total electricity consumption \cite{building_DOE} and still remains, to a large extent, unleveraged for distribution grid operations. The primary obstacle is in the current inability to accurately aggregate and synchronously operate a large ensemble of such small-scale loads, while taking into account their inherent techno- and socio-economic characteristics (e.g., dispatch limits, complex thermodynamics of building environments, and/or comfort preferences of building occupants). Therefore, to address these challenges, this paper focuses on mathematical modeling of an ensemble of thermostatically controlled loads (TCL), such as  heat pumps, air conditioners, heating and ventilation systems, for its accurate representation in energy management (dispatch) tools used by DR aggregators or local electric power utilities, \cite{aggregated_tcl_callaway,aggregated_tcl_lu}. 

The primary challenge in modeling TCL ensembles is to simultaneously achieve a high level of accuracy and maintain computational tractability. Currently, there are two large groups of methods to model and
forecast electricity  consumption of TCL ensembles: (i) physics-based co-simulation of TCLs and building dynamics (e.g. using heat transport models, electromechanical considerations, Kirchoff’s laws, evaporation, etc) and (ii) data-driven (e.g. statistical analyses and inference). The advantage of using the physics-based models is in their ability to describe buildings without prior observations. However, the performance of these models is highly sensitive to the number and accuracy of the
underlying modeling choices and assumptions, as well as to input parameters. Physics-based models often require more inputs than existing data acquisition systems can provide \cite{Coaklet_data_review}, and therefore incur significant uncertainties in both model parameters and dynamic processes. Using such models for controlling an ensemble of TCLs may lead to computational issues that
would prevent their scalability and implementation for real-life decision-making. On the other hand, in lieu of the physics-based models,
one can use machine learning and statistical modeling to perform data-driven studies of TCL and building  dynamics using a vast amount of historical data available at the buildings equipped with smart meters. These reduced order models are trained using the historical energy consumption data and other parameters (e.g. weather conditions, daily operational schedules, and control functionality) \cite{Koch_DR,CHASSIN_DR}. This paper develops a data-driven model to accurately represent a TCL ensemble using historical data and to continuously improve the accuracy of model performance via learning. 

\begin{figure}[!t]
\centering
\includegraphics[trim={2cm 3.7cm 2cm 2cm},clip,width=\columnwidth]{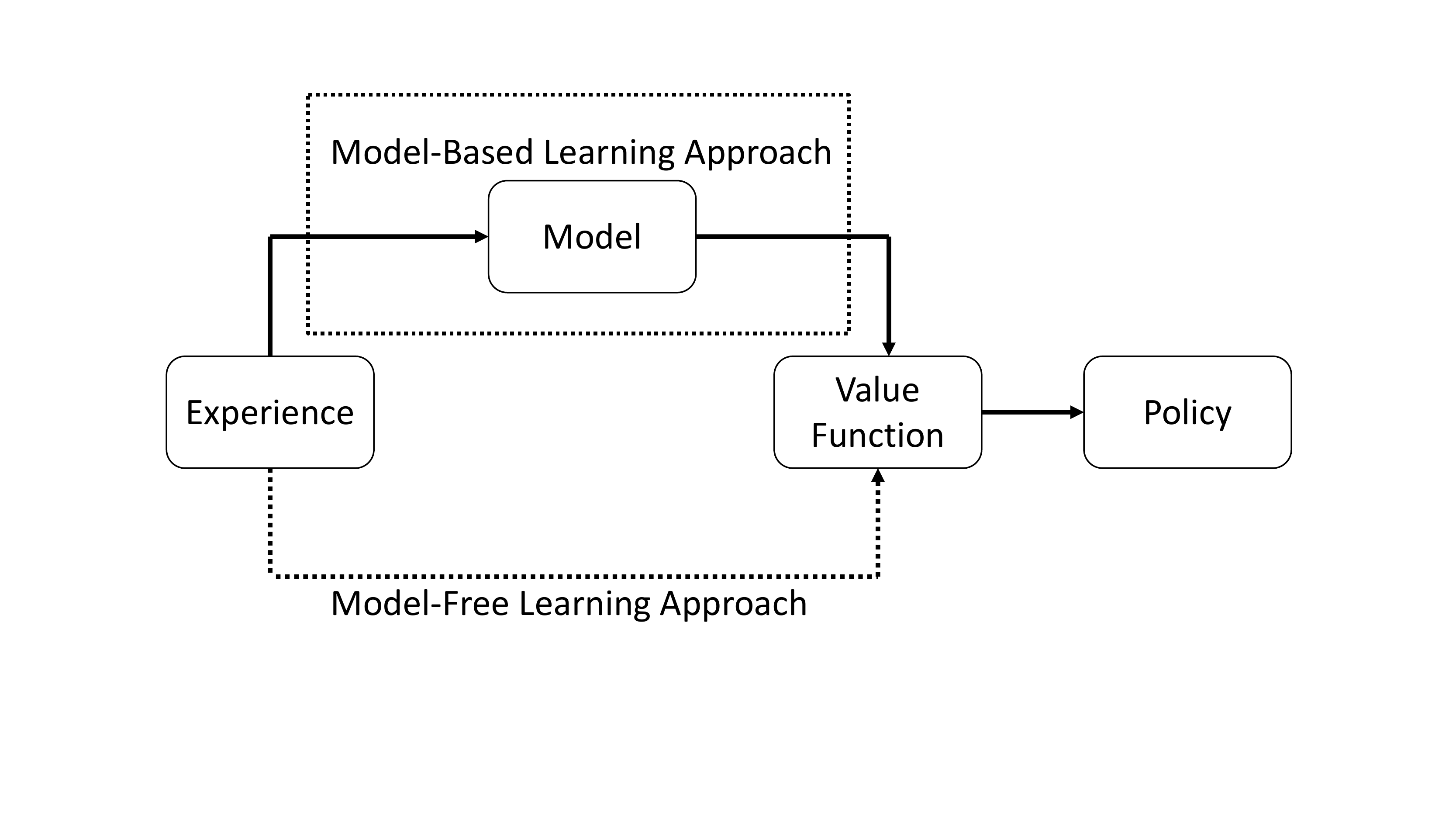}
\vspace{-20pt}
\caption{Comparison between the model-based and model-free learning approaches.}
\label{fig:block_fig}
\end{figure}

Among data-driven methods, TCL ensembles have been modelled as virtual storage units with linear dynamics, \cite{callawaystorage,mathieustorage,bowenstorage}, or as a Markov Decision Process (MDP) with probabilistic transitions, \cite{MDP_Mathieu,Meyn_MDP,Chertkov_MDP_chap,MDP_buildings,MDP_Network,MDP_Network_CC}. The MDP framework is particularly suitable for modeling large TCL ensembles, without sacrificing modeling accuracy or computational tractability. Thus, it produces high-quality solutions by means of using dynamic programming, which  are both analytically and computationally tractable. The models in \cite{MDP_Mathieu,Meyn_MDP,Chertkov_MDP_chap,MDP_buildings,MDP_Network,MDP_Network_CC} model a TCL ensemble as a discrete-time, discrete-space Markov Process characterized by a given transition probability matrix with deterministic coefficients. However, in practice, it is hardly possible to estimate these coefficients accurately due to the imperfection or incompleteness of historical measurements and behavioral uncertainty of consumers.  Therefore, the common caveat  of current MDP models in \cite{MDP_Mathieu,Meyn_MDP,Chertkov_MDP_chap,MDP_buildings,MDP_Network,MDP_Network_CC} is that they ignore uncertainty on model parameters (e.g. transition probabilities).  Since the inaccuracies stemming from the inability to compute model parameters in the MDP framework can be significant and can eliminate the benefits of using these resources for DR flexibility, this paper enhances the MDP framework with  model-free reinforcement learning (RL), where the DR aggregator\footnote{Alternatively, TCL ensembles can be aggregated and operated by utilities.} interacts with the TCL ensemble and  learns model parameters from both historical and streaming data (see Fig.~\ref{fig:block_fig}). \textcolor{black}{The main advantage of the model-free RL in the context of dispatch TCLs is in its ability to eliminate the need for knowing precise model parameters (e.g. parameters of the transition probability distribution underlying the MDP) because the optimal control policy can be learned  from \enquote{experience}.  In the context  of real-life DR applications, this \enquote{experience} can be obtained via indirect (passive) observations of the TCL ensemble or, in some cases, even individual TCLs by means of using advanced  metering infrastructure or data crowdsourcing, \cite{8233189}. }

Although there is a number of model-free RL techniques that can be used under the MDP framework, we exploit the property of TCL ensembles that allow for reducing a conventional MDP to a linearly-solvable MDP (LS-MDP). This reduction assumes that devices in  the TCL ensemble are relatively heterogeneous and, therefore, explicit control actions on each TCL device (e.g. on/off decisions or power consumption) can be replaced by a distribution of potential future states of the TCL ensemble, \cite{Todorov,optimal_actions_Todorov,LS_optimal_control}. Thus,  the optimal policy derived from the LS-MDP is not a mapping of states to action variables, as in a conventional MDP,  but is a mapping of a current state into a next-state distribution, which minimizes the expected next-state costs and the divergence cost between the default (e.g., without external control applied) and controlled   (e.g. with external control applied) probability distributions \cite{Hierarchical_LSMDP,LS_optimal_control}. The reduced LS-MDP problem is suitable for the Z-learning method, which is a modification of the common Q-learning method. In turn, the Z-learning method is capable of producing an accurate approximation of the original MDP at a faster convergence rate than the Q-learning method, \cite{Todorov,optimal_actions_Todorov,Hierarchical_LSMDP,LS_optimal_control}, mainly because Z-learning does not require state-action values as needed in Q-learning.

This paper uses a LS-MDP to model a  TCL ensemble and leverage the Z-learning method to find the optimal TCL dispatch policy. The Z-learning  method samples transitions passively from the default (e.g. without external control) behavior of the system, but is able to learn the optimal policy by leveraging the specific structure of LS-MDP. Note that the available state transitions may not accurately reflect the underlying true distribution due to limited availability of data. \textcolor{black}{Hence, we show that the Z-learning algorithm is robust to noise in the observed transitions and analyze its convergence in cases with and without noise.} The case study is carried out on aggregated heating, ventilation, and air conditioning (HVAC) systems in a residential neighborhood with 100 homes, where data is sampled using  the Net-Zero Energy Test Facility \cite{Nist_data}, operated by the \textcolor{black}{National Institute of Standards and Technology (NIST).}

The remainder of this paper is organized as follows. Section II presents a LS-MDP model for optimally dispatching a given  TCL ensemble. Section III solves the LS-MDP model using dynamic programming and leverages the  Z-learning approach to improve the solution accuracy. Section IV presents the case study using real-life data from the NIST Test Facility to demonstrate the usefulness of the proposed approach. Section V concludes the paper.

\section{Formulation}

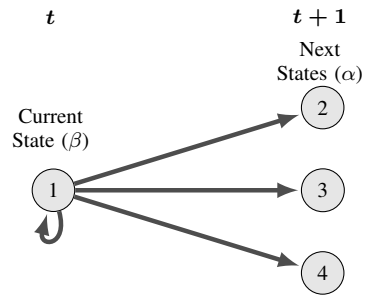
\begin{figure}[!b]
\centering
\vspace{-6mm}
  \begin{tikzpicture}[font=\footnotesize]
  \node[state,draw=black,fill=gray!20!white,minimum size=0.3cm] (s1) {1};
  \node[state,fill=gray!20!white,
     right=3.0cm of s1,yshift=1.1cm,minimum size=0.5cm] (s2) {2};
  \node[state,fill=gray!20!white,
     right=3.0cm of s1,yshift=0cm,minimum size=0.5cm] (s3) {3};
  \node[state,fill=gray!20!white,
     right=3.0cm of s1,yshift=-1.1cm,minimum size=0.5cm] (s4) {4};
  \node[right=-1.2cm of s1,yshift=0.8cm,font=\footnotesize,text width=1.5cm,align=center] {Current State ($\beta$)};
  \node[right=-1.2cm of s1,yshift=2.3cm,font=\footnotesize,text width=1.5cm,align=center] {$\boldsymbol{t}$};
  \node[right=-1.2cm of s2,yshift=0.6cm,font=\footnotesize,text width=1.5cm,align=center] {Next States ($\alpha$)};
  \node[right=-1.2cm of s2,yshift=1.2cm,font=\footnotesize,text width=1.5cm,align=center] {$\boldsymbol{t+1}$};
  \draw[>=latex,every loop,fill=black!70,
    draw=black!70,
     auto=right,
     line width=0.4mm]
    (s1) edge[line width=0.6mm] (s2)
    (s1) edge[line width=0.6mm] (s3)
    (s1) edge[line width=0.6mm] (s4)
    (s1) edge[loop below,line width=0.6mm]  (s1);
  \end{tikzpicture}
  \vspace{-1mm}
  \caption{A schematic representation of the Markov Process displaying transitions from a current state ($\beta$) at time $t$ to the possible future next states ($\alpha$) at time $t+1$. Note that the ensemble can remain in the same state $\beta$ at time $t+1$ such that $\alpha$ = $\beta$.}
  \label{mdp:states}
\end{figure}
\textcolor{black}{Similarly to \cite{Chertkov_MDP_chap,MDP_buildings,MDP_Network,MDP_Network_CC}, the MDP framework is leveraged to build the model for the control of the TCL ensemble.} We define a LS-MDP for modeling a given TCL ensemble  as a 5-tuple $\{ \mathcal{T}$, $\mathcal{A}$, $U_t^{\beta}$, $\mathcal{P}_t^{\alpha\beta}$, $\overline{\mathcal{P}}^{\alpha\beta}\}$,
where $\mathcal{T}$ is the set of time intervals, which constitute a planning horizon, $\mathcal{A}$ is the set of possible states, \textcolor{black}{$U_t^\beta$ is the utility of the aggregator} in state $\beta \in \mathcal A$ at time $t \in \mathcal{T}$,  $\overline{\mathcal{P}}^{\alpha\beta}$ and ${\mathcal{P}}_{t}^{\alpha\beta}$ are  default (i.e. without control actions of the DR aggregator) and controlled (with  control actions of the DR aggregator) transition probabilities from state $\beta \in \mathcal A$ to  $\alpha \in \mathcal A$. The states in set $\mathcal{A} = \big\{ \alpha, \beta, ...\big\}$ are obtained by discretizing the range of power consumption for each TCL ensemble given the operating range of TCL devices in the \textcolor{black}{ensemble}. For any given state $\beta \in \mathcal{A}$ at time $t\in\mathcal{T}$, the probability of the transition of the TCL ensemble to the next state $\alpha \in \mathcal{A}$ at time $t+1\in\mathcal{T}$ is characterized by $\mathcal{P}^{\alpha\beta}_{t}$. Fig.~\ref{mdp:states} displays all possible transitions from the current state $\beta$ at time $t$ to all possible next states $\alpha$ at time $t+1$. Note that the ensemble can remain in the same state $\beta$ at time $t+1$ such that $\alpha$ = $\beta$. The default transition probabilities, represented by parameter $\mathcal{\overline{P}}^{\alpha\beta}$, corresponds to the internal dynamics of the TCL ensemble without actions of the aggregator and are typically estimated from historical data (see \cite{MDP_buildings}). The TCL ensemble is then optimized as:
\begin{subequations}
\begin{align}
&\underset{\substack{\rho,\mathcal{P}}}{\text{min}} \ \mathbb{E}_{\rho}
\sum_{t \in \mathcal{T}-1}  \sum_{\alpha \in \mathcal{A}} \big(-U_{t+1}^{\alpha} + \sum_{\beta \in \mathcal{A}} \gamma \log\! \frac{\mathcal{P}_{t}^{\alpha\beta}}{\overline{\mathcal{P}}^{\alpha\beta}}\big)  \label{MDP:obj}
\\
&\rho_{t+1}^{\alpha} = \sum_{\beta \in \mathcal{A}} \mathcal{P}_{t}^{\alpha\beta} \rho_{t}^{\beta}, \quad \forall \alpha \in \mathcal{A}, t \in \mathcal{T} 
\label{MDP_evol} \\
&\sum_{\alpha \in \mathcal{A}} \mathcal{P}_{t}^{\alpha \beta} = 1,\quad \forall \beta \in \mathcal{A},  t \in \mathcal{T}-1, \label{mdp_integrality}  
\end{align}
\label{TCL_optimization}
\end{subequations}
where $\rho_{t+1}^{\alpha} \geq 0$ and $\rho_{t}^{\beta} \geq 0$ are decision variables, which characterize the probability that the TCL ensemble is operated in states $\alpha$ and $\beta$ at time $t+1$ and $t$, and are related via  transition probabilities $\mathcal{P}_{t}^{\alpha\beta}$. Eq.~\eqref{MDP:obj} represents the objective function of the DR aggregator that controls the TCL ensemble and aims to maximize its expected utility \textcolor{black}{or minimize its expected cost of energy} ($-U_{t+1}^{\alpha}$) and to minimize the discomfort cost for the TCL ensemble. The discomfort cost is computed using the Kullback-Leibler (KL) divergence, weighted by parameter $\gamma$. This divergence penalizes the difference between the transition decisions made by the DR aggregator ($\mathcal{P}_{t}^{\alpha\beta}$) and the default transitions of the TCL ensemble ($\mathcal{\overline{P}}^{\alpha \beta}$), under the assumption that the latter represents first-choice preferences of TCL users. \textcolor{black}{Parameter $\gamma$ can influence the KL divergence and thus encourage or discourage deviations from the default behavior of the TCL ensemble.} The choice of the KL  divergence for the penalty cost is motivated by its extensive use for modeling randomness of discrete and continuous time-series \cite{KL_divg}. Eq.~\eqref{MDP_evol} describes the temporal evolution of the TCL ensemble from time $t$ to $t+1$ over time horizon $\mathcal{T}$. Eq.~\eqref{mdp_integrality} imposes the integrality constraint on the transition decisions optimized by the DR aggregator such that their total probability is equal to one. 

\textcolor{black}{After solving \eqref{TCL_optimization} as described later in Section \ref{Solving LS-MDP}, the active power ($p_{t}$) consumed by the TCL ensemble can be computed  using optimized decisions $\rho^{\beta}_{t}$ and rated active power $p^{\beta,rated}$ at each state, e.g.  $p_{t}= \sum_{\beta \in \mathcal{A}} p^{\beta,rated} \rho_{t}^{\beta},\forall t \in \mathcal{T}$.} 


\subsection{\textcolor{black}{Relation to Other Methods}}
The LS-MDP in \eqref{TCL_optimization} can be related to linear dynamical TCL models in other data-driven methods, \cite{callawaystorage,mathieustorage}. \textcolor{black}{Consider the following linear dynamics for the TCL ensemble, \cite{bowenstorage}:}
\begin{align}
    S_{t+1}=S_{t}+ (u_t+P_t)\Delta t,\label{LTImodel}
\end{align}
where $S_t$ is the energy state of the TCL ensemble, $P_t$ is the normal power consumed and $u_t$ is the power change sought by control actions. Let $P_t \sim N(\mu_{P_t}, \sigma_{P_t})$ and consider the KL divergence between  $S^0_{t+1}$ (without control) and $S_{t+1}$ (with control). Using \cite{bishop2006pattern} leads to: 
\begin{align}
    KL(S^0_{t+1}||S_{t+1}) = \frac{u^2(t)}{2\sigma^2_{P}\Delta^2t},
\end{align}
which is  the quadratic cost for control used in linear systems, i.e. the quadratic discomfort cost for control in linear dynamics with Gaussian uncertainties is equivalent to the discrete-time KL cost in the LS-MDP. However, the discrete nature of LS-MDP transitions simplifies modeling, even for complex state transitions and non-Gaussian uncertainties.

\section{Z-learning in LS-MDP} \label{Z-learning in LS-MDP}
\subsection{Solving LS-MDP} \label{Solving LS-MDP}
The optimization in Eq.~\eqref{TCL_optimization} is a Linearly Solvable MDP (LS-MDP) as introduced by \cite{Todorov}. The  optimal policy for Eq.~\eqref{TCL_optimization} is computed using techniques from dynamic programming \cite{RL}. The Bellman equation for the LS-MDP in \eqref{bellmen_1} can be derived from the Bellman equation for the traditional MDP  explained in Appendix \ref{bellman_LSMDP_Appendix} and leads to:
\begin{align}
\begin{split}
& \frac{1}{\gamma}\varphi^{\beta}_{t} = \frac{1}{\gamma} \underset{\substack{\mathcal{P}}}{\text{min}} \Big(-U_{t}^{\beta} + \mathbb{E}_{\mathcal{P}^{\alpha\beta}_{t}} \! \Big[\gamma \log \frac{\mathcal{P}_{t}^{\alpha\beta}} {{\overline{\mathcal{P}}^{\alpha\beta}}} + \varphi^{\alpha}_{t+1}\Big] \Big), \label{bellmen_1}  
\end{split} 
\end{align}

where $\varphi^{\beta}_{t}$ is the value function of the TCL ensemble at  present state $\beta$ at time $t$ and $\varphi^{\alpha}_{t+1}$ is the value function at  next state $\alpha$ at time $t+1$. By introducing desirability function $z^{\beta}_{t} = \text{exp}(\frac{-\varphi^{\beta}_{t}}{\gamma})$ in \eqref{bellmen_1} we obtain:

\begin{align}
\begin{split}
& -\!\!\text{log}(\!z^{\beta}_{t})\! =\! \frac{1}{\gamma} \underset{\substack{\mathcal{P}}}{\text{min}} \Big(\!\!-\!U_{t}^{\beta}\!\! +\!  \gamma  \mathbb{E}_{\mathcal{P}^{\alpha\beta}_{t}}\! \Big[\! \log\! \frac{\mathcal{P}_{t}^{\alpha\beta}} {{\overline{\mathcal{P}}^{\alpha\beta}}} \!-\! \text{log}(\!z^{\alpha}_{t+1})\! \Big] \Big) = 
\end{split}\nonumber\\
\begin{split}
&\frac{1}{\gamma} \underset{\substack{\mathcal{P}}}{\text{min}} \bigg(-U_{t}^{\beta} + \gamma  \mathbb{E}_{\mathcal{P}^{\alpha\beta}_{t}} \bigg[ \log \frac{\mathcal{P}_{t}^{\alpha\beta}} {{\overline{\mathcal{P}}^{\alpha\beta}}z^{\alpha}_{t+1}} \bigg]  \bigg) \label{bellmen_5}
\end{split}
\end{align}
After introducing  a normalization term defined as $\mathcal{G}^{\beta}_{t}(z)=\sum_{\alpha}\overline{\mathcal{P}}^{\alpha\beta}z^{\alpha}_{t+1}$,  \eqref{bellmen_5} can be recast as:

\begin{align}
\begin{split}
& \!-\text{log}(z^{\beta}_{t})\! = \!\frac{1}{\gamma} \underset{\substack{\mathcal{P}}}{\text{min}}\!  \bigg(\!\!\!-U_{t}^{\beta}\! +\! \gamma  \mathbb{E}_{\mathcal{P}^{\alpha\beta}_{t}}\! \bigg[\! \log  \frac{\mathcal{P}_{t}^{\alpha\beta}\mathcal{G}^{\beta}_{t}(z)} {{\overline{\mathcal{P}}^{\alpha\beta}}z^{\alpha}_{t+1} \mathcal{G}^{\beta}_{t}(z)} \!\bigg]\! \bigg)\\& = \bigg(\frac{-U_{t}^{\beta}}{\gamma} + \underset{\substack{\mathcal{P}}}{\text{min}} 
KL\! \bigg[\mathcal{P}_{t}^{\alpha\beta} \bigg\Vert \frac{\overline{\mathcal{P}}^{\alpha \beta}z^{\alpha}_{t+1}}{\mathcal{G}^{\beta}_{t}(z)}  \bigg]  - \text{log}\mathcal{G}^{\beta}_{t}(z)\! \bigg) \label{bellmen_9} 
\end{split}
\end{align}
The KL divergence provides the expectation of the log-difference between the two distributions such that \textcolor{black}{ $KL[p_{1}\Vert p_{2}] = \mathbb{E}_{p_1}[\log\frac{p_1}{p_2}]$.} It is zero if and only if the two distributions are same. Therefore, it follows from Eq.~\eqref{bellmen_9} that the optimal policy is achieved when  the KL divergence term in Eq.~\eqref{bellmen_9} is minimal, i.e. it is equal to zero. Hence, by equating the two distributions in the KL divergence, the optimal policy follows as:
\begin{align}
&\mathcal{P}_{t}^{\alpha \beta} = \frac{\overline{\mathcal{P}}^{\alpha \beta}z^{\alpha}_{t+1}}{\mathcal{G}^{\beta}_{t}(z)}  = \frac{\overline{\mathcal{P}}^{\alpha \beta}z^{\alpha}_{t+1}}{\sum_{\alpha}\overline{\mathcal{P}}^{\alpha \beta}z^{\alpha}_{t+1}}, \label{optimal_policy_1}
\end{align}
The optimal policy in Eq.~\eqref{optimal_policy_1} depends on the uncontrolled transition probability ($\overline{\mathcal{P}}^{\alpha \beta}$) and the desirability function of the TCL ensemble at the next state ($z^{\alpha}_{t+1}$). The optimal policy reduces the Bellman equation in \eqref{bellmen_9} to the following form:
\begin{align}
\begin{split}
& -\text{log}(z^{\beta}_{t}) = \{\frac{-U_{t}^{\beta}}{\gamma} -\text{log}\mathcal{G}^{\beta}_{t}(z) \} \label{bellmen_reduced_1} 
\end{split} \\   
\begin{split}
& \text{log}(z^{\beta}_{t}) =  \Big\{\frac{U_{t}^{\beta}}{\gamma}  + \text{log} \Big[\sum_{\alpha}\overline{\mathcal{P}}^{\alpha \beta}z^{\alpha}_{t+1} \Big]  \Big\} \label{bellmen_reduced_2} 
\end{split}
\end{align}
Exponentiating Eq.~\eqref{bellmen_reduced_2} converts the Bellman equation to the following reduced form:
\begin{align}
&z^{\beta}_{t} = \text{exp}\Big(\frac{U_{t}^{\beta}}{\gamma}\Big) \sum_{\alpha}\overline{\mathcal{P}}^{\alpha \beta}z^{\alpha}_{t+1} \label{bellmen_reduced_3}
\end{align}
Given the Bellman equation in \eqref{bellmen_reduced_3} and optimal policy in \eqref{optimal_policy_1}, the LS-MDP is solved as described in Algorithm \ref{alg:lsmdp}. \textcolor{black}{Eq.~\eqref{bellmen_reduced_3} is linear and thus can be represented in a matrix form as $\boldsymbol{z}_t = \boldsymbol{U}_t \overline{\mathcal{P}} \boldsymbol{z}_{t+1}$, where $\boldsymbol{z}_t$ is a vector with elements $z^{\beta}_{t}$, $\overline{\mathcal{P}}$ is a matrix with entries $\overline{\mathcal{P}}^{\alpha\beta}$, and $\boldsymbol{U}_t$ is a diagonal matrix with elements $\text{exp}\Big(\frac{U_{t}^{\beta}}{\gamma}\Big)$ along its main diagonal \cite{optimal_actions_Todorov,Todorov,LS_optimal_control,Hierarchical_LSMDP,Busic_TAC,Busic_CDC}.}
\begin{algorithm}[!t]
\caption{Solving a LS-MDP}
\label{alg:lsmdp}
\begin{algorithmic}[1]
\State Initialize $z^{\beta}_{t}=1$ $\forall \beta \in \mathcal{A}$, $t\in\mathcal{T}$ 
\State $z^{\beta}_{|\mathcal{T}|} = \text{exp}\Big(\frac{U_{|\mathcal{T}|}^{\beta}}{\gamma}\Big)$, where $|\mathcal{T}|$:=final time 
\State Set $t=|\mathcal{T}-1|$
\For{$t\gets |\mathcal{T}-1|$ \textbf{to} $1$} 
    \State $z^{\beta}_{t} = \text{exp}\Big(\frac{U_{t}^{\beta}}{\gamma}\Big) \sum_{\alpha}\overline{\mathcal{P}}^{\alpha \beta}z^{\alpha}_{t+1}$
\EndFor
\For{all $t\in\mathcal{T}-1$}{ 
    \For{all $\beta\in\mathcal{A}$} 
    \State $\mathcal{P}_{t}^{\alpha \beta} = \frac{\overline{\mathcal{P}}^{\alpha \beta}z^{\alpha}_{t+1}}{\sum_{\alpha}\overline{\mathcal{P}}^{\alpha \beta}z^{\alpha}_{t+1}}$
    \EndFor
\EndFor}
\end{algorithmic}
\end{algorithm}

\subsection{Z-learning}
Although the LS-MDP solves the optimization problem for the TCL ensemble efficiently, it requires knowledge about the model of the environment. Since the model is estimated from the historical data (e.g. values of the default transitions in $\overline{\mathcal{P}}^{\alpha\beta}$), which is limited and imperfect, it may introduce inaccuracies. This motivates the use of model-free learning techniques to robustly solve the optimization problem in \eqref{TCL_optimization}.  Using Z-learning, a model-free learning method, returns stochastic approximations  $\hat{z}$ of the optimal value function in Eq.~\eqref{bellmen_reduced_3}. Thus, $\hat{z}$ is updated as 
\begin{align}\label{Z-updateeq}
&\hat{z}^{\beta}_{t,k} \leftarrow (1-\eta_k)\hat{z}^{\beta}_{t,k-1} + \eta_k \text{exp} \bigg(\frac{U^{\beta}_{t}}{\gamma} \bigg) \hat{z}^{\alpha}_{t+1,k-1}
\end{align}
where $\eta_k$ is a decaying learning rate and $\alpha$ is the state observed at sample $k$ by transitioning from previous state $\beta$. Z-learning updates the value function at the present state based on the sample providing next-state information instead of averaging over all the future possible states as in the LS-MDP. Unlike in Q-learning, there is no optimization of actions during the iterations in Z-learning. Instead, the samples for  Z-learning are passively collected from the underlying distribution discretized in  $\overline{\mathcal{P}}^{\alpha\beta}$. Then, $\hat{z}^{\beta}_{t,k}$ are updated by using the specific KL divergence form of the optimal policy, which enables faster computations.

\textcolor{black}{The proposed application of the Z-learning algorithm to dispatching TCLs  is detailed in Algorithm \ref{alg:zlearning}. First, the algorithm is initialized with $z_t^{\beta}=1$ for all states  $\beta \in \mathcal{A}$ and time periods $t \in \mathcal{T}$. Next, it computes  the desirability function for the final time $|\mathcal{T}|$. Then, it iteratively computes  the desirability function for the remaining time intervals (from $t=1$ to $t=|\mathcal{T}-1|$) using  samples generated from the passive dynamics and updates the desirability function until a chosen convergence criterion  is achieved. In this paper, the convergence criterion is defined as the difference between two successive values of the desirability function.}

\begin{algorithm}[!t]
\caption{Z-learning}
\label{alg:zlearning}
\begin{algorithmic}[1]
\State Initialize $z^{\beta}_{t}=1$ $\forall \beta \in \mathcal{A}$, $t\in\mathcal{T}$ 
\State $z^{\beta}_{|\mathcal{T}|} = \text{exp}\Big(\frac{U_{|\mathcal{T}|}^{\beta}}{\gamma}\Big)$, where $|\mathcal{T}|$:=final time 
\Repeat
    \State Set $k$ = current sample at state $\alpha$ from passive dynamics $\overline{\mathcal{P}}$
    
    \State Starting with time $t=|\mathcal{T}-1|$
    \For{$t\gets |\mathcal{T}-1|$ \textbf{to} $1$}
        \State $\hat{z}^{\beta}_{t,k} \leftarrow (1-\eta_k)\hat{z}^{\beta}_{t,k-1} + \eta_k \text{exp} \bigg(\frac{U^{\beta}_{t}}{\gamma} \bigg) \hat{z}^{\alpha}_{t+1,k-1}$
        \EndFor
\Until{convergence}
\end{algorithmic}
\end{algorithm}
Note that the state transitions in the samples used in Z-learning may be corrupted by noise as well. The noise in the passive dynamics is modelled as the error term $\boldsymbol{\epsilon}^{\alpha\beta} \in \mathbb{R}^{n\text{x}n}$, where $n=|\mathcal{A}|$ :
\begin{equation}
\boldsymbol{\overline{\mathcal{P}}}^{\alpha \beta} = \overline{\mathcal{P}}^{\alpha \beta} + \boldsymbol{\epsilon}^{\alpha\beta} \label{z_noise}
\end{equation}
where $\boldsymbol{\epsilon}^{\alpha\beta}$ can be modelled by a zero-mean, normal distribution with variance $\sigma_{n}^{2}$, i.e. $\boldsymbol{\epsilon}^{\alpha\beta}\sim N(0,\sigma^{2})$ (other parametric distributions are also suitable). To ensure that every row in the transition probability matrix remains equal to one i.e. $\sum_{\alpha \in \mathcal{A}} \boldsymbol{\overline{\mathcal{P}}}^{\alpha\beta} = 1, \forall \beta \in \mathcal{A}$, every row in $\boldsymbol{\epsilon}^{\alpha\beta}$ must be equal to zero, i.e. $\sum_{\alpha \in \mathcal{A}} \boldsymbol{\epsilon}^{\alpha\beta} = 0, \forall \beta \in \mathcal{A}$ \footnote{\textcolor{black}{Other methods can be used to capture noise, such as Interval Markov Chains, where actual transition probabilities lie in intervals \cite{Interval_MC}.}}. $\boldsymbol{\overline{\mathcal{P}}}^{\alpha \beta}$ can be extended to capture noise scenarios by defining a set of $N$ probability distributions as $\boldsymbol{\overline{\mathcal{P}}}_{n}^{\alpha \beta}, \forall n \in [1,N]$, such that  $\boldsymbol{\overline{\mathcal{P}}}_{n}^{\alpha \beta}$ is characterized as:
\begin{align}
& \frac{1}{N} (\mathbb{E}[\boldsymbol{\overline{\mathcal{P}}}_{1}^{\alpha \beta}] + \mathbb{E}[\boldsymbol{\overline{\mathcal{P}}}_{2}^{\alpha \beta}] + \dots + \mathbb{E}[\boldsymbol{\overline{\mathcal{P}}}_{N}^{\alpha \beta})] \approx \overline{\mathcal{P}}^{\alpha \beta}, \label{noise_N_prob} 
\end{align}
where Eq.~\eqref{noise_N_prob} ensures that the expected value of all $N$ distributions is close to the passive dynamics of the TCL ensemble given by $\overline{\mathcal{P}}^{\alpha \beta}$. At each Z-learning iteration, one out of $N$ distributions is selected with $\frac{1}{N}$ probability to update the value function. Note that despite the noise in the transition probability matrix, the same Algorithm \ref{alg:zlearning} for Z-learning is used and, as shown in Section~\ref{sec:casestudy}, performs efficiently and robustly.

\subsection{Convergence of Z-learning} The convergence of Z-learning can be assessed using the optimal LS-MDP policy in \eqref{optimal_policy_1} by proving that the Z-update in \eqref{Z-updateeq} asymptotically converges  to  \eqref{bellmen_reduced_3}. Let $\Delta \hat{z}_{t,k}^\beta = z^\beta_t-z_{t,k}^\beta$ be the optimality at  the $k^{th}$ iteration of Z-learning. Using \eqref{bellmen_reduced_3}-\eqref{Z-updateeq}  leads to:
\begin{align*}
   \Delta \hat{z}_{t,k}^\beta = &(1-\eta_k)\Delta\hat{z}_{t,k-1}^\beta \nonumber\\
   +&\eta_k\text{exp} \bigg(\frac{U_{t}^{\beta}}{\gamma}\bigg)(\mathbb{E}_{\overline{\mathcal{P}}}[z^{\alpha}_{t+1} ]- \sum_\alpha \mathbbm{1}_{\alpha_k = \alpha}\hat{z}_{t+1,k-1}^\alpha),
\end{align*}
where the indicator function $\mathbbm{1}$ is  $1$, if state $\alpha$ is observed in the $k^{th}$ iteration, and $0$ otherwise.  Consider $t = |\mathcal{T}|-1$, the final time-interval for updating $z$-values. $\hat{z}_\mathcal{T} =z_\mathcal{T}$ is of course directly determined using $\mathcal{U}_\mathcal{T}$. It is clear that  $\mathbb{E}_{\overline{\mathcal{P}}}[z^{\alpha}_{\mathcal T}]-\sum_\alpha \mathbbm{1}_{\alpha_k = \alpha}z_{\mathcal{T},k}^\alpha$ is a random variable with mean $0$ and a finite variance. Then, if learning rates $\eta_k$ are selected such that $\sum_k \eta_k =\infty$ and $\sum_k \eta^2_k <\infty$,  it follows that $\lim_k \Delta \hat{z}_{{\mathcal T}-1,k}^\beta\rightarrow 0$, see \cite{jaakkola1994convergence}. Following similarly for $t = |\mathcal{T}|-2, \ldots, 1$, it returns $\lim_k \Delta \hat{z}_{t,k}^\beta \rightarrow 0$. Thus Z-update \eqref{Z-updateeq} converges to the solution of \eqref{bellmen_reduced_3}. Note that the convergence also holds if a finite variance noise is allowed in the transition probability matrix (see Eq.~(\ref{z_noise})).


\section{Case Study} \label{sec:casestudy}

\subsection{Data}
\begin{figure}[!t]
\centering
\includegraphics[width=\columnwidth]{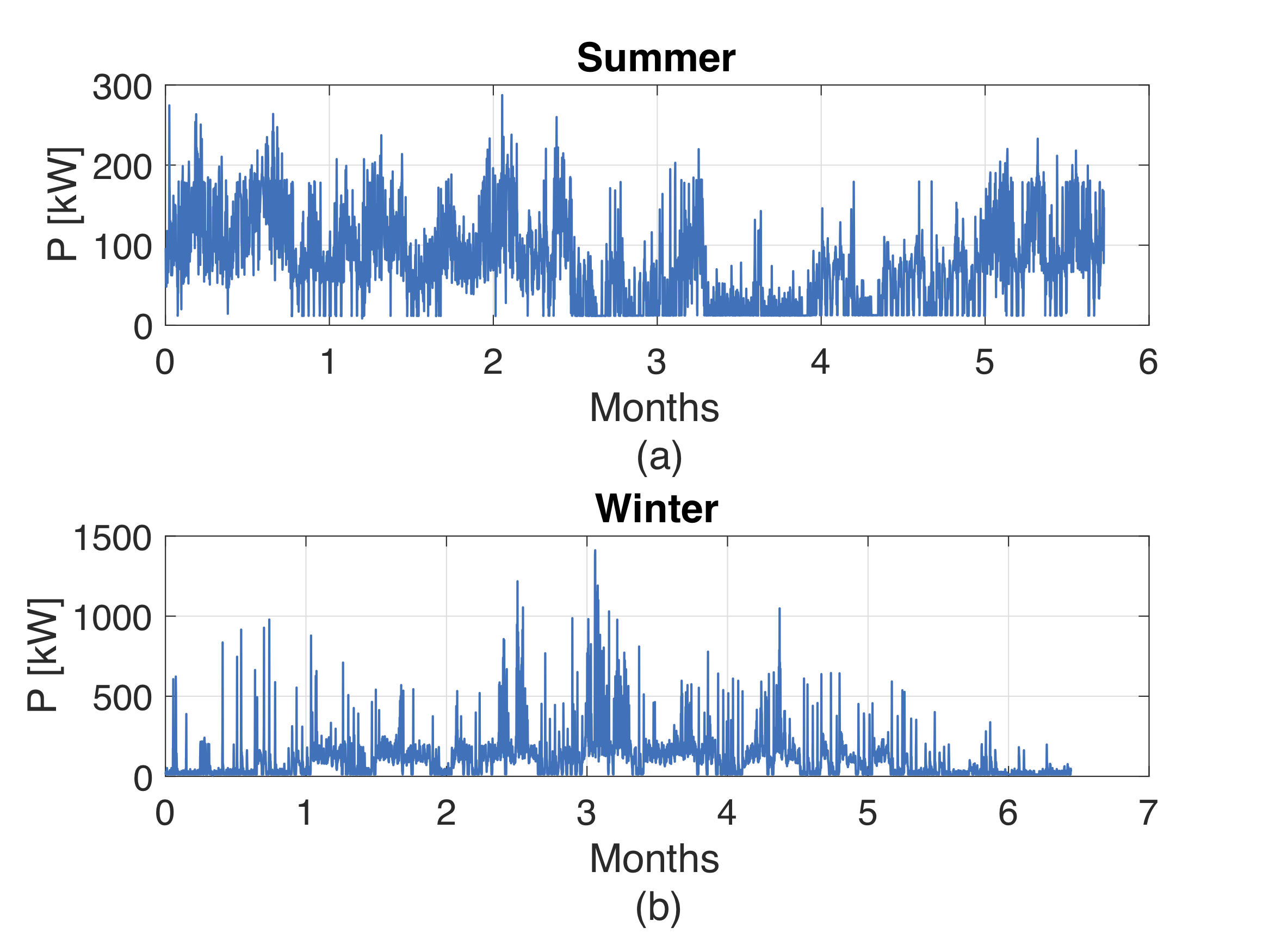}
\vspace{-20pt}
\caption{Aggregated HVAC power consumption of 100 houses.}
\label{fig:nist_power}
\end{figure}

\begin{figure}[!t]
\centering
\includegraphics[width=\columnwidth]{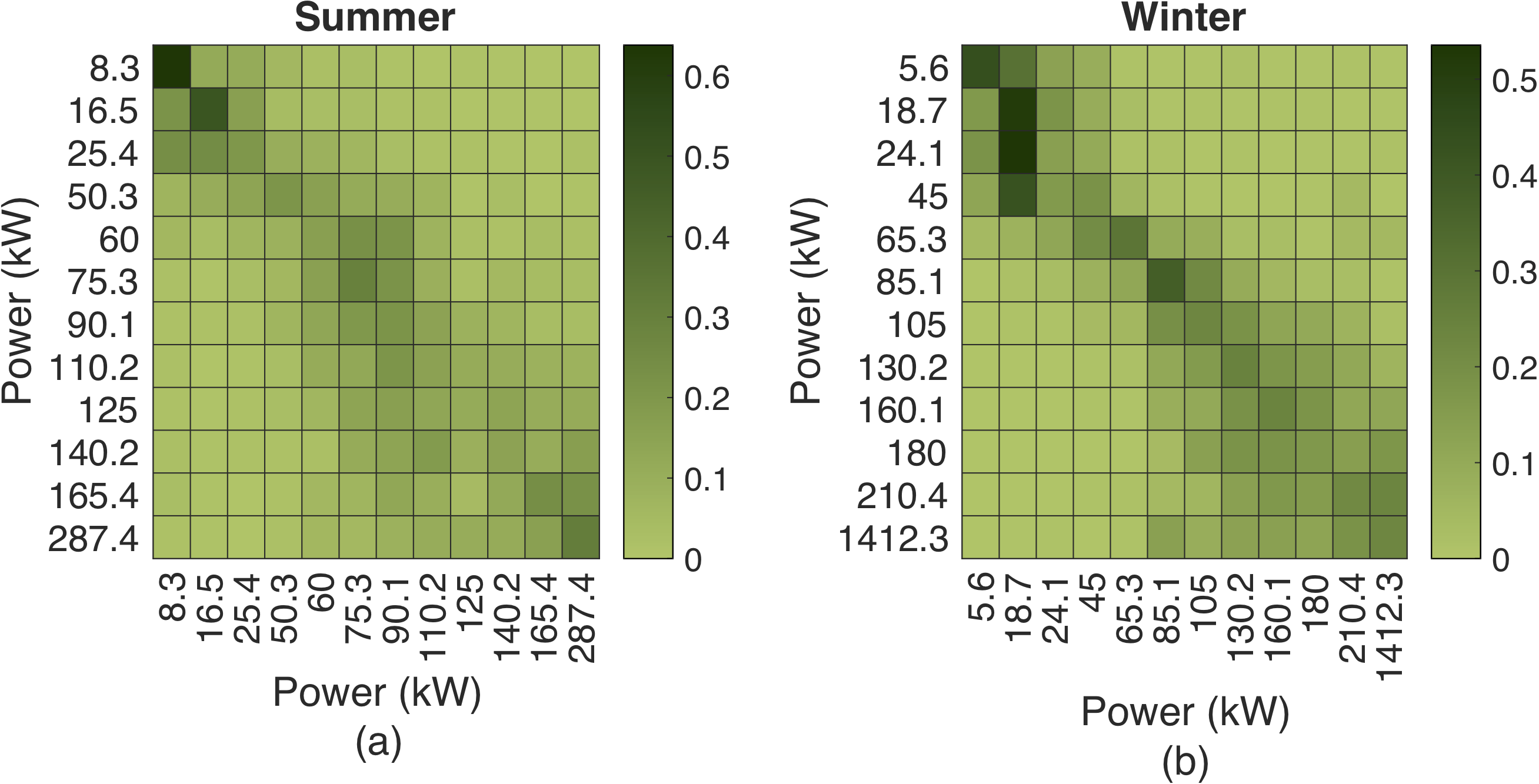}
\vspace{-20pt}
\caption{Default transition probability matrix with 12 states constructed from the power profiles in Fig.~\ref{fig:nist_power}, where color density indicates the probability value in the sidebar.}
\label{fig:states}
\end{figure}
We use data from the  Net-Zero Energy Test Facility, \cite{Nist_data}, which is a single-family, three-floor, net-zero-energy house, with the total area of 386 (4156) m$^2$ (ft$^2$), located  in Gaithersburg, MD. To create an ensemble, this case study considers  a neighbourhood with 100 houses with parameters and historical data obtained based on adding random noise to the data obtained from the Net-Zero Energy Test Facility. The random noise is limited in its magnitude by 20\% of the original values because 100 houses are assumed to be located in close proximity and function similarly. For these 100 houses, we extract HVAC data and assume that all HVACs are operated by the same DR aggregator. Fig \ref{fig:nist_power} shows the aggregated HVAC power consumption for both summer and winter seasons in the period from July 1, 2013 to June 30, 2014. These profiles were discretized in 12 Markovian states  and Fig. \ref{fig:states} displays the resulting transition probability matrices ($\overline{\mathcal{P}}^{\alpha\beta}$). \textcolor{black}{These transition probability matrices are used to dispatch the TCL ensemble over the time horizon of 10 hourly intervals.}

The case study solves the TCL optimization problem in Eq.~\eqref{TCL_optimization} using Algorithms \ref{alg:lsmdp} and \ref{alg:zlearning} and compare their performance in terms of the value function using the  error metric:

\begin{equation}
\text{Error} = \frac{\sum_{\beta\in\mathcal{A}}|\varphi^{\beta_{\text{LS-MDP}}}_{t}-\varphi^{\beta_{\text{Z-learning}}}_{t}|}{\sum_{\beta\in\mathcal{A}}(\varphi^{\beta_{\text{LS-MDP}}}_{t})} \label{eq:error},
\end{equation}
which computes the relative difference between the  Z-learning and LS-MDP values. Moreover, the Z-learning algorithm is run for two cases: (a) without noise added to the passive dynamics and (b) with noise. \textcolor{black}{The learning rate for the Z-learning algorithms is set to decay as $\eta_k=\frac{1000}{1000+k}$, where $k$ is a sample number.}

\subsection{Results}
\begin{figure}[!t]
\centering
\includegraphics[width=\columnwidth]{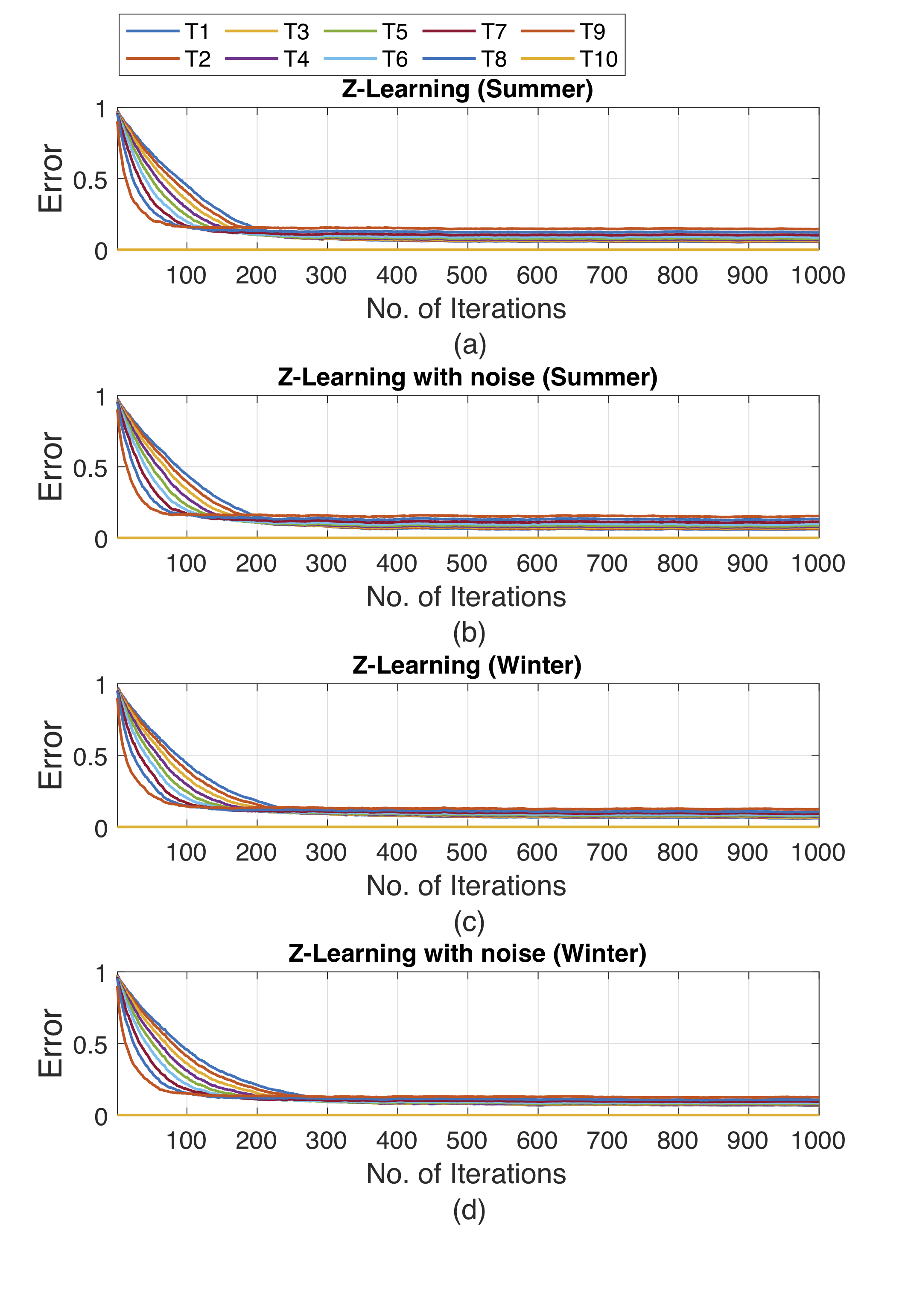}
\vspace{-40pt}
\caption{Comparing Z-learning performance with and without noise during the summer and winter seasons \textcolor{black}{for hourly time periods T1-T10.}}
\label{fig:err_both}
\end{figure}

Fig.~\ref{fig:err_both} describes the error convergence of the Z-learning algorithm with and without noise for each hourly time period.  As the number of learning iterations increases, the resulting error reduces. The rate of convergence differs for the winter and summer seasons. For instance, the 10\% error for all time period is achieved within 225 and 245 learning iterations. Similarly, the effect of noise on the learning rate is more visible during the winter season, where the number of learning iterations required to achieve the 10\% error  increases from 245 to 290 iterations. In contrast, in the summer case, adding noise does not affect the convergence rate and Z-learning achieve the 10\% error in 225 learning iterations. The slower convergence rate in the winter case is explained by the fact that a greater power consumption being approximated using the same number of discrete states in the transition probability matrix, which requires more  exploration of the model environment, especially when noise samples noticeably deviate from the default behavior defined by the passive dynamics.

\begin{figure}[!t]
\centering
\includegraphics[width=\columnwidth]{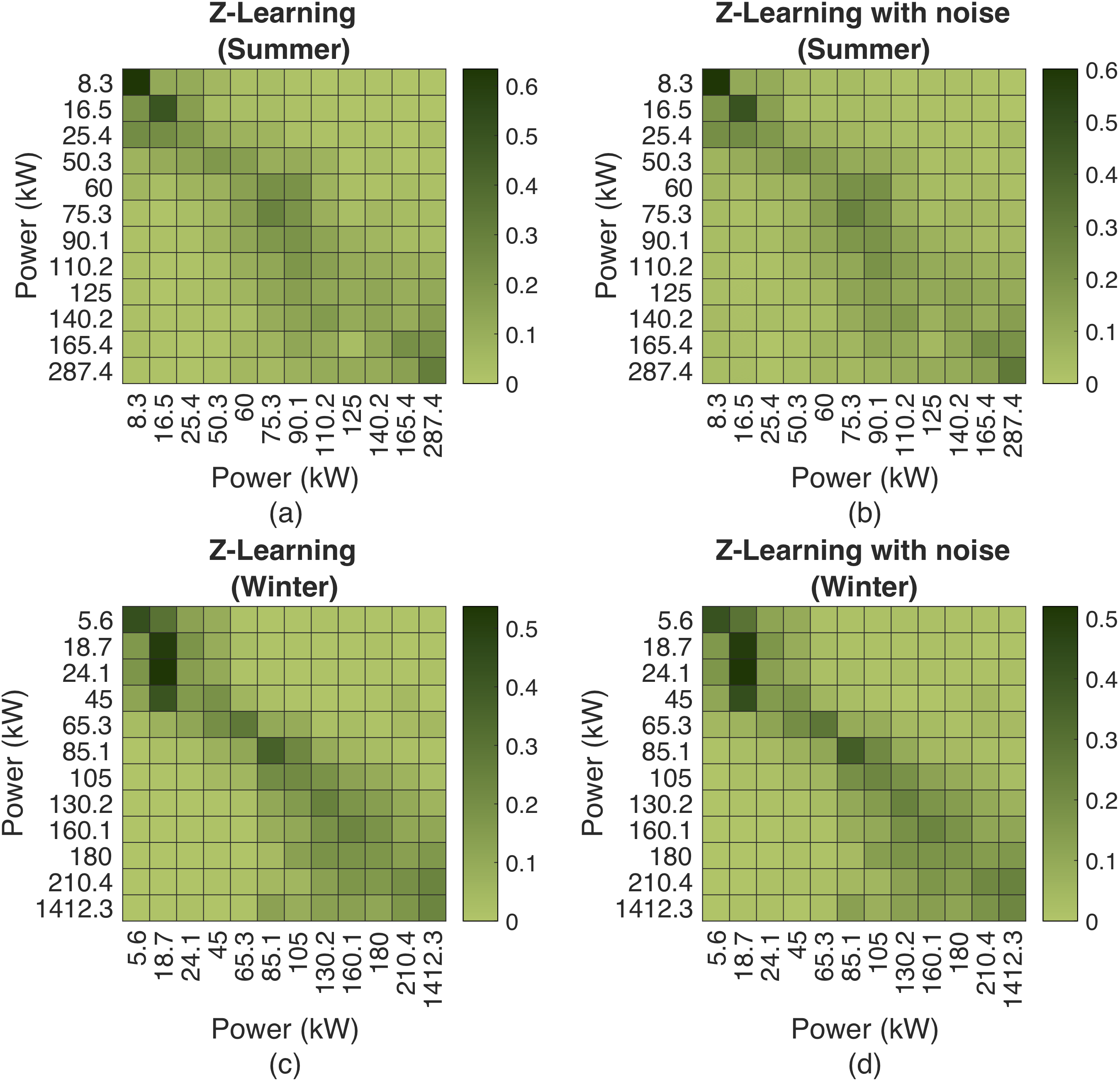}
\vspace{-17pt}
\caption{Estimated transition probabilities for the summer and winter seasons with and without noise.}
\label{estimated_prob_both}
\vspace{-5mm}
\end{figure}

Given the outcomes of  Z-learning, the estimated transition probabilities are obtained as shown in Fig.~\ref{estimated_prob_both}. The estimated matrices for the cases with and without noise do not differ significantly. Thus, the Root-mean-square difference of elements between is 0.0101\% and 0.0068\% for the summer and winter seasons. Notably, this difference changes only slightly when compared to the default transition matrices in Fig.~\ref{mdp:states}. In the case of winter season shown in Fig.~\ref{estimated_prob_both} (c) and (d), the difference is 0.0017\% and 0.0055\% for the case without noise and with noise. The difference for the summer season in Fig.~\ref{estimated_prob_both} (a) and (b) increases relative to the winter season and is 0.0023\% and 0.01\% for the case without noise and with noise. The result of using  Z-learning is that as the number of iterations and samples increases, its outcomes will converge to the LS-MDP values. 

\begin{figure}[!t]
\centering
\includegraphics[width=\columnwidth]{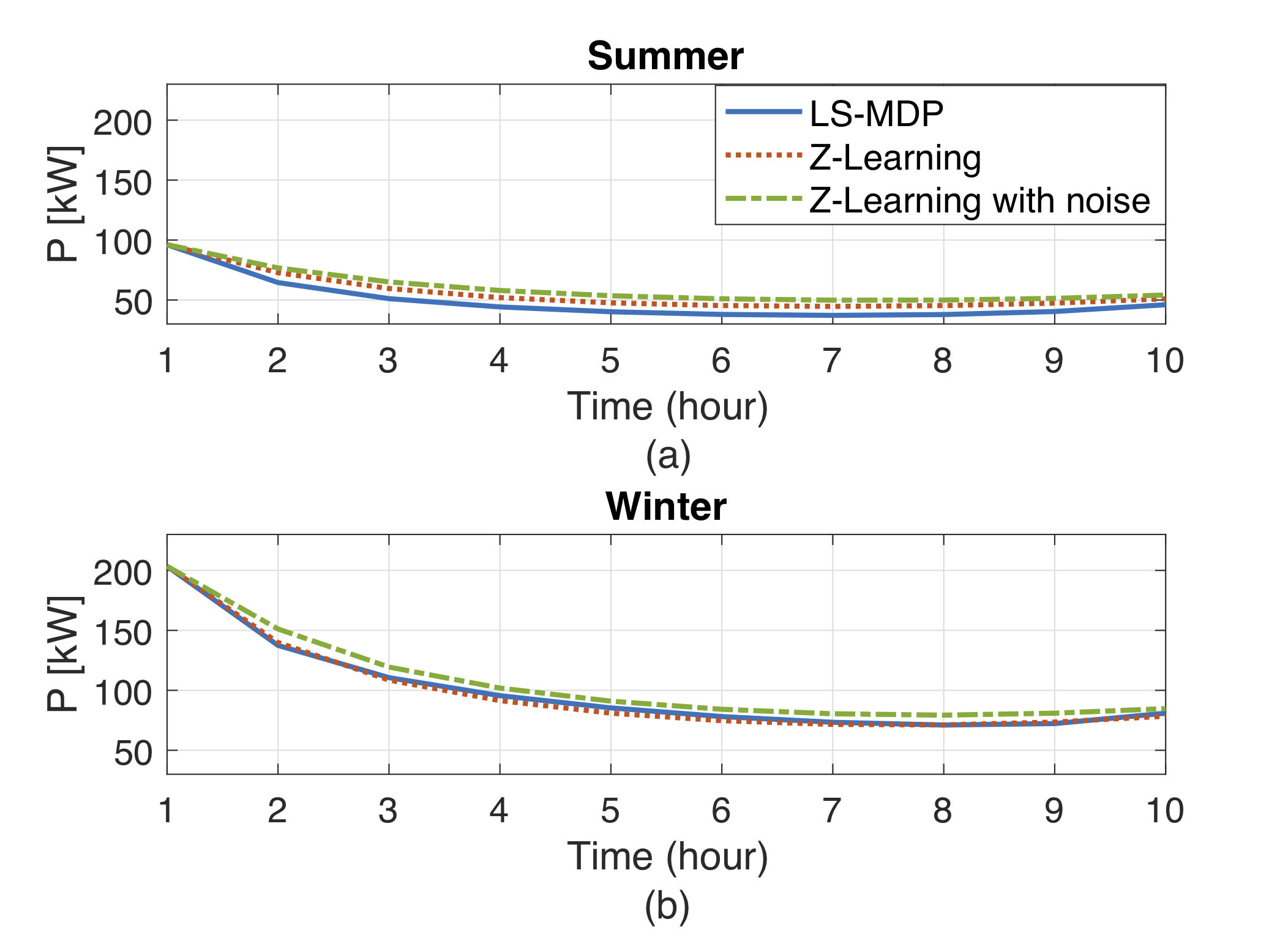}
\vspace{-20pt}
\caption{Comparison of the TCL ensemble dispatch decisions for the LS-MDP and Z-learning solutions.}
\label{fig:injections}
\end{figure}

\begin{figure}[!t]
\centering
\includegraphics[width=\columnwidth]{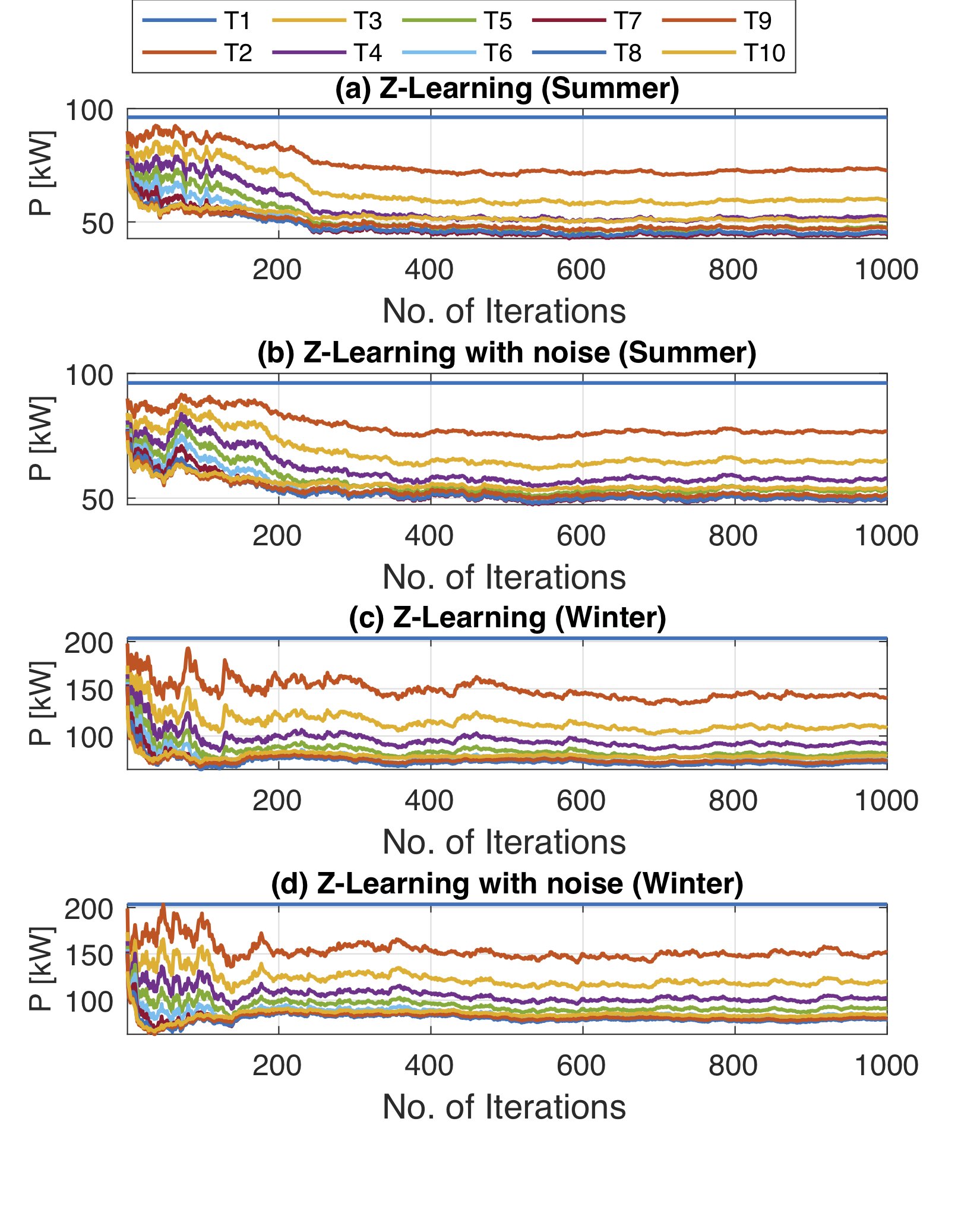}
\vspace{-30pt}
\caption{Dispatch decisions for the TCL ensemble obtained with Z-learning \textcolor{black}{for hourly time periods T1-T10.}}
\label{fig:power_itr}
\end{figure}

\begin{figure}[!t]
\centering
\includegraphics[width=\columnwidth]{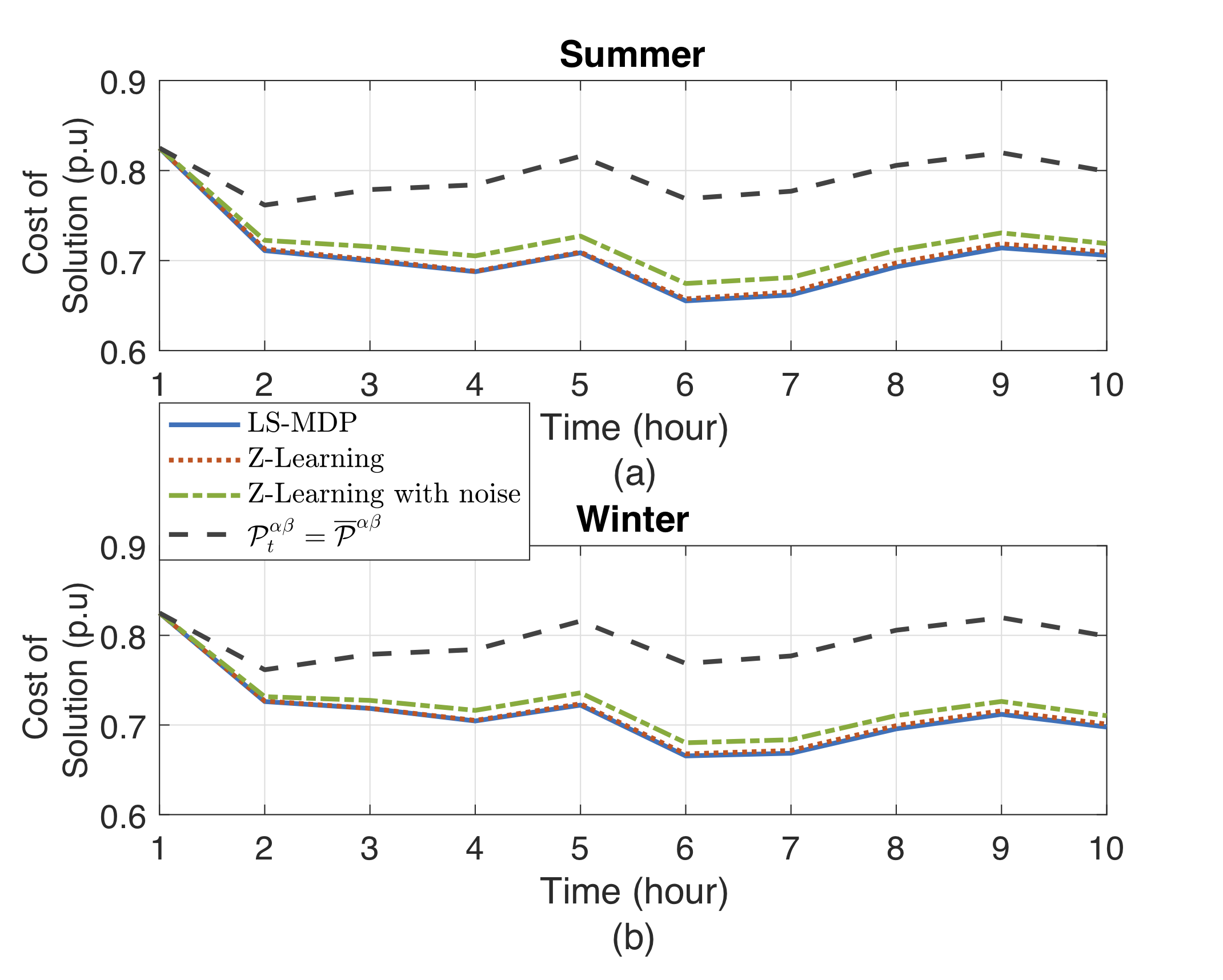}
\vspace{-20pt}
\caption{Comparison of the solution cost for the LS-MDP and Z-learning solutions.}
\label{fig:cost}
\end{figure}

Based on the transition probability matrices obtained with the LS-MDP and Z-learning, Fig.~\ref{fig:injections} compares the power dispatch of the TCL ensemble.  Both Z-learning results with and without noise accurately approximate the benchmark LS-MDP solution. The maximum difference observed for the case with noise is 4.17 kW for the summer season and 13.7 kW for the winter season, and for the case without noise is 13.7 kW for the summer season and 13.9 kW for the winter season. These differences are relatively small given the summer and winter peaks of 287.4 kW and 1412.3 kW. The power dispatch of the TCL ensemble at every iteration during Z-learning is shown in Fig.~\ref{fig:power_itr}, where values stabilize as the number of iterations continues to increase. Similarly, Fig.~\ref{fig:cost} compares the value function of each method that represents the operating cost of the TCL ensemble. The values of the operating cost for both Z-learning with and without noise are slightly greater than the optimal value provided by the LS-MDP, because Z-learning approximates the optimal solution for the optimization problem that minimizes the objective function (i.e. the operating cost). \textcolor{black}{Notably, the operating cost is comparatively high when $\mathcal{P}_t^{\alpha\beta} = \overline{\mathcal{P}}^{\alpha\beta}$ (no control taken), which shows the importance of controlling the TCL ensemble to lower the cost.}

\section{Conclusion}

This paper presents a data-driven learning method for the control of TCL ensemble using the MDP and Z-learning approaches. The results show the importance of moving from model-based methods to model-free methods to bridge the gap between real environment and its model. The importance of modelling uncertainty to provide more robust solutions is demonstrated by comparing the TCL ensemble injections and cost of the solution. In future, we will also consider the related problem of TCL optimization under uncertain energy prices and analyze the regret associated with online learning based schemes \cite{neu2017fast}. 

\appendices
\section{Bellman Equation Derivation for LS-MDP} \label{bellman_LSMDP_Appendix}
The Bellman equation for a  finite-horizon MDP is \cite{bellma_mdp}:
\begin{align}
&\frac{1}{\gamma}\varphi^{\beta}_{t} = \frac{1}{\gamma} \underset{u}{\text{min}}\{l^{\beta}_{t}(u) + \mathbb{E}_{{\mathcal{P}}^{\alpha \beta}_{t}(u)}[\varphi^{\alpha}_{t+1}] \}, \label{bellmen}
\end{align}
where $l^{\beta}_{t}(u)$ represents the immediate cost that the agent pays at time $t$ for taking action $u$ at state $\beta$ and $\mathbb{E}_{\mathcal{P}^{\alpha \beta}_{t}(u)}[\varphi^{\alpha}_{t+1}]$ is the expectation of $\varphi^{\alpha}_{t+1}$ taken with respect to ${\mathcal{P}}^{\alpha \beta}_{t}(u)$: 
\begin{align}
&\mathbb{E}_{{\mathcal{P}}^{\alpha \beta}_{t}(u)}[\varphi^{\alpha}_{t+1}] = \sum_{\alpha} {{\mathcal{P}}^{\alpha \beta}_{t}(u)}\varphi^{\alpha}_{t+1}, 
\end{align}
Eq.~\eqref{bellmen} implicate the search over all actions $u$ for each new state $\alpha$. However, this can be time consuming due to the exponential growth of future states. The LS-MDP offers a solution for this problem, which uses the transition probabilities instead of the symbolic actions, where the agent can directly specify the probability of transition from the current state to any possible future state. The Bellman equation for choosing  $\mathcal{P}^{\alpha \beta}_{t}$ by the agent  is:
\begin{align}
\begin{split} &\frac{1}{\gamma}\varphi^{\beta}_{t} = \frac{1}{\gamma} \underset{\mathcal{P}}{\text{min}}\Big\{ U^{\beta}_{t} \! +\! \gamma \mathbb{E}_{\mathcal{P}^{\alpha\beta}_{t} }\bigg[ \text{log}\frac{\mathcal{P}^{\alpha \beta}_{t}}{{\overline{\mathcal{P}}^{\alpha \beta}}} \bigg] \!+\! \mathbb{E}_{{\mathcal{P}}^{\alpha \beta}_{t}}[\varphi^{\alpha}_{t+1}] \Big\},\label{bellmen_linear}
\end{split}
\end{align}
where $U^{\beta}_{t}$ represents the state cost and $\mathbb{E}_{\mathcal{P}^{\alpha\beta}_{t}}$ means the statistical expectation of $\alpha$ taken with respect to the controlled transition distribution ${\mathcal{P}^{\alpha\beta}_{t}}$. Eq.~\eqref{bellmen_linear} represents the Bellman equation for LS-MDP.

\bibliographystyle{IEEEtran}
\bibliography{ref.bib}

\end{document}